\documentclass[12pt]{article}
\usepackage{times}

\usepackage[cm]{fullpage}
\usepackage{multicol,graphicx,enumitem,eurosym}

\usepackage[utf8]{inputenc}
\usepackage{fullpage}
\usepackage{color,url,graphicx}

\newcommand{\ie}{\emph{i.e.~}}
\newcommand{\eg}{\emph{e.g.~}}
\newcommand{\etal}{\emph{et al.~}}

\title{\vspace*{-37pt}VigiFlood: evaluating the impact \\\vspace*{-0pt} of a change of perspective on flood vigilance}
\author{Carole Adam - \url{Carole.Adam@imag.fr}\\
\emph{Univ. Grenoble Alpes, LIG, Grenoble, F-38000 France}}
\date{May 2020}

\begin{document}
\maketitle

\begin{abstract}
Emergency managers receive communication training about the importance of being 'first, right and credible', and taking into account the psychology of their audience and their particular reasoning under stress and risk. But we believe that citizens should be similarly trained about how to deal with risk communication. In particular, such messages necessarily carry a part of uncertainty since most natural risks are difficult to accurately forecast ahead of time. Yet, citizens should keep trusting the emergency communicators even after they made forecasting errors in the past.

We have designed a serious game called Vigiflood, based on a real case study of flash floods hitting the South West of France in October 2018. In this game, the user changes perspective by taking the role of an emergency communicator, having to set the level of vigilance to alert the population, based on uncertain clues. Our hypothesis is that this change of perspective can improve the player's awareness and response to future flood vigilance announcements. We evaluated this game through an online survey where people were asked to answer a questionnaire about flood risk awareness and behavioural intentions before and after playing the game, in order to assess its impact. 

\end{abstract}

\section{Introduction}

In October 2018, the Aude department in the South-West of France was hit by intense rain over many hours. Flash floods were hard to forecast and only the 'orange' level of vigilance could be raised initially, but the population dismissed this very usual warning for that season. The 'red' level was then raised too late during the following night, after several towns were already flooded and some people had died. This led to high criticism from the population and in the media following the events. The main problem here is the loss of trust induced by too many 'false alarms'. 

Indeed, emergency managers receive risk communication training: they learn the importance of being 'first, right and credible', which is far from easy in an unpredictable context with high stakes; they are also taught how to take into account the psychology of their audience and their particular reasoning under stress and risk. But there is no real training for citizens about how to deal with risk communication. They are not told that such messages necessarily carry a part of uncertainty since most natural risks are difficult to accurately forecast ahead of time (e.g. earthquake, flash flood, etc). As a result, they lose trust after what they interpret as 'false alerts', and decrease their vigilance during subsequent events, which negatively impacts their safety. The population should therefore be made (more) aware of the fact that 'false alerts' are difficult to avoid, without risking on the contrary to miss events, which could have serious consequences; and that they are responsible for being vigilant. We believe that a good way to achieve better awareness is to get them to change perspective by taking on the other role, and ''playing'' the role of an emergency communicator.

To address this issue, we propose a serious game called \textbf{VigiFlood} for raising awareness in the population about the difficulty of crisis communication and their own responsibility for reacting to the alerts. The player has to take on the role of an emergency communicator setting the level of vigilance based on uncertain weather forecast. They get feedback about the effect of their actions on the population (trust, evacuation decision).
We expect this change of perspective to let the players gain awareness about the difficulty of risk prediction and communication, and subsequently lead to a change in behaviour. The game is based on a multi-agent model grounded on psychological and sociological theories, and on real hydrological and meteorological data from the geographical area. A first version is implemented, but improvements are still ongoing. Our hypothesis is that such a role-playing game, by getting the user to change perspective, will improve their awareness.
We evaluated the validity of this hypothesis via an online questionnaire simulating an extract of the game, preceded and followed by questions assessing the responder's awareness. 

This paper is structured as follows. Section~\ref{sec:floods} introduces the context: flash floods, the French vigilance system, and our case study. Section~\ref{sec:soa} provides the reader with an overview of related literature in different areas: crisis communication, trust and risk theories, agent-based models of natural disasters, and serious games. Section~\ref{sec:vigiflood} describes Vigiflood, the underlying agent-based model, the implementation, and the game principles. Section~\ref{sec:eval} is the core contribution and presents our evaluation of Vigiflood, the questionnaire design, its ethical validation, and the results obtained from an analysis of 80 answers gathered during the month of June 2019. Finally, Section~\ref{sec:cci} concludes and discusses future prospects of this research.

%%%%%%%%%%%%%%%%%%%%%%%%%%%%%%
\section{Context} \label{sec:floods}

\subsection{Flash floods}
Flash floods generally occur due to rapid rain on an already saturated soil (after a particularly wet period) or on a soil with a poor absorption capability (such as concrete, as is often the case in urbanised settings), or due to extensive rain because of a storm or hurricane. They can also occur from more occasional events such as a glacier melting after a volcanic eruption, an ice dam melting, or a man-made dam failing. In this paper we focus on rain-induced flash floods, whose prediction depends on the meteorological services. 

The indicators used by forecast services include: forecast quantity of rain (radar or satellite or model based), soil absorption capacity, soil moisture or dryness level, topography, basin or catchment capacity, etc \cite{cepri}. 

\subsection{French flood vigilance system}

In France, the agency in charge of forecasting the expected amount of rain is Meteo France\footnote{\url{http://www.meteofrance.com/previsions-meteo-france/previsions-pluie}}. Weak rain is defined as 1 to 3 mm/h, moderate rain is between 4 and 7 mm/h, and heavy rain is over 8 mm/h. Another agency is in charge of monitoring the main waterways and broadcast their expected height and debit 3 to 6 hours ahead of time: Vigicrues\footnote{\url{www.vigicrues.gouv.fr}}. However, only part of the waterways are monitored, and the European Flood Risk Prevision Center (CEPRI\footnote{\url{https://www.cepri.net/}} notices that half of the 63 victims in the Languedoc-Roussillon region (south of France) alone between 1996 and 2006 died on catchment basins that were not monitored. 

\begin{figure}[ht]
    \centering
    \includegraphics[scale=0.27]{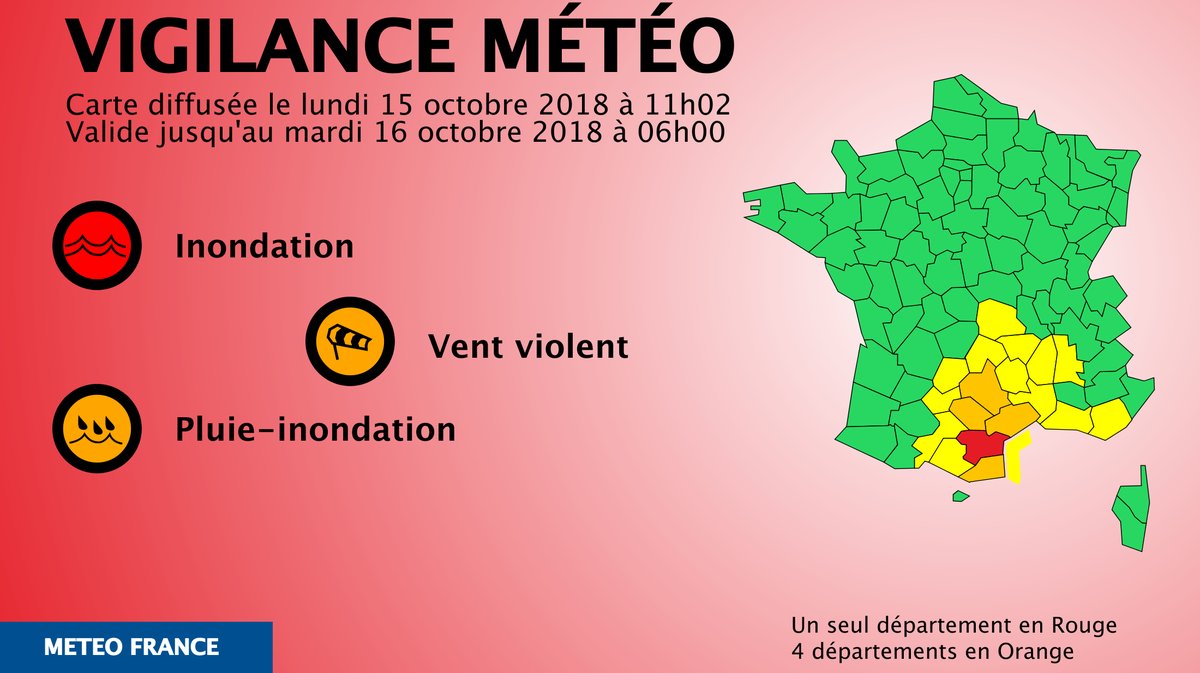}
    \caption{MeteoFrance vigilance map, 15 October, 11am (source: MeteoFrance)}
    \label{fig:vigimf}
\end{figure}

The meteorological services in charge of the area then analyses these clues (rain forecast, waterway height and debit, etc) to define and announce a level of vigilance on a 4-colour scale, from green (no problem), yellow, orange, to red (higher risk). This level of vigilance is publicly available online\footnote{See: \url{http://vigilance-public.meteo.fr/} or: \url{http://vigilance.meteofrance.com/}} (example map on Figure~\ref{fig:vigimf}). Each region has its particularities and can be more or less used to receiving heavy rain in a short amount of time, so that the vigilance thresholds are not the same everywhere.

\subsection{Case study: Aude floods, October 2018}

\begin{figure}[ht]
    \centering
    \includegraphics{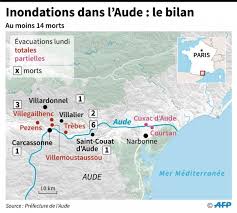}
    \caption{Map of the flooded area, with victim counts, and evacuated towns. Source: Préfecture de l'Aude}
    \label{fig:map}
\end{figure}

The Aude department in France (see map on Figure~\ref{fig:map}) depends on the CNP center in Toulouse; its critical rain threshold for the orange vigilance is 50-100 mm in 24 hours, with a regional record of 551 mm in 24 hours. On Sunday 14 October 2018 the meteorological station of Carcassonne received 139.8 mm of rain. However the vigilance level was first only set to orange, until Monday 15 October morning at 6am when it was finally raised to red. In the meantime, several towns were already flooded, roads and bridges destroyed, and some people had died. The final toll of these floods in the Aude department is 15 dead and 99 injured people, 204 towns classified as hit by ``natural disaster’’, and a total cost of damages estimated to 220 M\euro.

As a result, MeteoFrance vigilance system was harshly criticised in the media \cite{libe2018}. The representative of the Ministry of the Interior, Frédéric de Lanouvelle, interviewed on LCI-TV, evoked a \emph{``weakness in the orange vigilance level which is used very often and when there is a real problem, people do not take it into account anymore''} (our translation). He adds that based on residents' statements, the red vigilance level was indeed raised too late, but explains it is due to the difficulty in forecasting such a powerful episode. 

However, as noted by the Major Risks Institute (IRMA) \cite{irmavigi2018}, there had only been 3 orange vigilance raised in 2018 in the Aude department for rain-floods before October 14: January 7-8; 28 February and 1st of March; and the last one on October 9-10, immediately following an orange vigilance for thunderstorms on October 8-9. Besides, this last episode probably influenced the dramatic floods of October 14-15 by saturating the soil with water. \cite{irmavigi2018} therefore concludes that the general public must be made aware of their responsibilities, and be taught that several orange vigilances in a row should make them particularly vigilant, instead of the opposite, dismissing the last vigilance because nothing serious happened during the previous ones.

\section{Related work} \label{sec:soa}

This section provides a quick overview of relevant literature about crisis communication, risk evaluation, trust; indeed, these works will ground our conceptual model of the population. It also reviews some existing simulations and games applied to disaster awareness.

%%%%%%%%%%%%%%%%%%%%%%%%%%%%%%%%%
\subsection{Crisis communication}
Reynolds (\cite{reynolds2010principles}) identifies 6 principles of efficient crisis emergency and risk communication (CERC):
\begin{itemize}
    \item \textbf{Be First}: in times of crisis, people need information fast, and will often stick with the first source they get information from. 
    \item \textbf{Be Right}: information should be accurate and transparent, explaining what is known or not, and what is being done to get more information.
    \item \textbf{Be Credible}: communicators should be honest and trustful, and avoid to promise what they cannot deliver, at the risk of losing the public's trust.
    \item \textbf{Express Empathy}: communicators should acknowledge the population feelings and challenges to build trust and a good relationship.
    \item \textbf{Promote Action}: messages should not only be informative but also action-oriented, to empower the population, give them a feeling of control, and release their anxiety.
    \item \textbf{Show Respect}: communicators should account for the vulnerability of people during crises, their messages should be respectful and promote cooperation and rapport.
\end{itemize}

\cite{barry2013literature} review the literature to explain the ``complex and unpredictable ways that individuals perceive risk'', and find that risk communication is influenced by socio-cultural, environmental and linguistic factors. Covello \etal (\cite{covello2001risk,covello2003best}) identify 4 factors of risk communication: 
\begin{itemize}
    \item \textbf{Risk perception} is subjective, influenced by socio-cultural and cognitive factors (\eg the individual feeling of agency or control).
    \item \textbf{Mental noise} created by stress and threat reduces the ability to process information: technical and scientific concepts must be ``translated [...] into understandable messages'', which should be visual and repeated. 
    \item \textbf{Negative dominance}: upset people focus more on negative (losses) than positive (gains) information, so communication must counter-balance negative messages with many more positive, action-oriented messages, and focus on progress rather than failures.
    \item \textbf{Trust determination}: upset people tend to not trust authority, so trust must be created well ahead of any crisis, through proactive community communication.
\end{itemize}

Further, \cite{hurnen1997effect} discuss the importance of the locus of control of individuals: if \textbf{external} (\ie they believe they have no control) they will feel helpless, and be less likely to take preventative action or react to warnings than if \textbf{internal} (\ie they believe they can do something). This is in line with the CERC principle to 'promote action' and give people something to do so that they feel (more) in control. Similarly, \cite{khan2012influences} discusses the lack of response of the population when the perceived risk is either too high (fatalism) or too low (``blasé effect''): in both cases, people feel powerless. He concludes by recommending to pay careful attention to the level of risk that is communicated to the population in the warnings to avoid undesirable reactions or inertia.

The American Center for Disease Control and Prevention issued a report about Crisis and Emergency Risk Communication \cite{cercpsy} focusing on specific psychological factors of information processing during a crisis, and how messages should be adapted subsequently:
\begin{itemize}
    \item People \textbf{simplify} messages, might not hear, not remember, or misinterpret them. Logical reasoning is impaired and decisions might rely more on habits, routines or imitation. Messages should thus be simple.
    \item People \textbf{hold} on to their current beliefs, and prefer trusted and familiar sources of information even if inaccurate and non-expert, to reputable experts who might disagree with each other. Messages should therefore come from credible sources.
    \item People try to \textbf{confirm} information before acting, via complementary information and additional opinions. Messages should therefore be consistent between the different channels.
    \item People believe the \textbf{first} message, and compare any further message with it; lack of information creates anxiety, speculation and rumors. Accurate messages should therefore be released as early as possible.
\end{itemize}
These findings are in line with research showing the importance of cognitive biases in disasters \cite{murata2015influence,arnaud2017role}.

%%%%%%%%%%%%%%%%%%%%%%%%%%%
\subsection{Trust and risk}
According to Slovic (\cite{slovic2016chap19}), risk management has become much more ``contentious'', with risk managers blaming the public for being irrational, and the public blaming the stakeholders for their poor management. In his view risk communication, aiming at aligning population and experts' perceptions with experts, has failed due to the lack of trust: ``if trust is lacking, no form or process of communication will be satisfactory'' (p. 410).

Slovic also explains that trust is ``fragile'', builds up slowly but can be destroyed instantly and is then hard or impossible to regain (p. 319). He provides several reasons for this asymmetry: 
\begin{itemize}
    \item Negative events are more visible than positive ones (one missed alarm stands out in many days of correct predictions);
    \item Negative events have more weight because they have lower probability and higher consequences (a flood is rarer than a ``normal day'' and can do serious damage);
    \item The media also tends to give more coverage to bad news than good news;
    \item Sources of bad news are seen as more credible, less likely to lie, than sources of good news;
    \item Distrust strengthens itself, by limiting further interactions and biasing future interpretations towards the reinforcement of existing (distrustful) beliefs.
\end{itemize}

%%%%%%%%%%%%%%%%%%%%%%%%%%%%%%%
\subsection{Agent-based simulations}
In this context, computer simulations can be a useful tool. Axelrod (\cite{axelrod1997advancing}) defines 7 purposes of simulations, including prediction (simulate a system very realistically to predict its future behaviour, \eg meteorology), training (provide a believable interactive environment to rehearse actions, \eg flight simulator), or education (let the user learn by trying in a virtual world). Predictive simulations require a high degree of realism to lay valid predictions, while training and education simulations can be simpler and less realistic (they ``need not be rich enough to suggest a complete real or imaginary world''). Besides, Axelrod also claims that ``the simpler the model, the easier it may be to discover and understand the subtle effects of its hypothesized mechanisms''; educational simulations are therefore often quite simple.

Simulations have often been used in crisis management for various purposes. For instance \cite{yang2018assessment} provide a very realistic model based on field data, to \textbf{predict} the impact of early warnings on population behaviour in terms of reducing material losses from floods. Others focus on realistically modelling the physical flood phenomenon in order to \textbf{support decisions} regarding early warnings for tsunamis \cite{friedemann2011explicit}, based on data from multiple sensors \cite{behrens2008handling}.

Other simulations focus on communication, but not necessarily during floods. \cite{arru2019} study if the population should be alerted or not of an ongoing crisis (\eg terrorist attack), depending on the anticipation of their potential reaction (\eg crowd panic), which is based on a psychological model. Since events considered are ongoing, they do not deal with false alarms and their potential impact on long-term trust. \cite{adam2018} study the propagation of awareness in the Australian population after a bushfires warning, depending on its channel and (familiar vs unfamiliar) source. Since they consider a single event, they do not deal with long-term dynamics of trust over multiple (right and wrong) alerts. \cite{adam2019} also proposed a serious game for trying various communication strategies to alert the population (focused vs wide targeting, information vs recommendations) before and during fires, but they do not deal with the timing of alerts nor the impact of false alerts. Indeed, they report no habituation phenomenon to fire alerts, which could be due to the easier predictability of fires compared to flash floods, or to normative and cultural differences. 

%%%%%%%%%%%%%%%%%%%%%%%%%%%%%%%%
%%% TODO TODO TODO TODO TODO %%%
%%%%%%%%%%%%%%%%%%%%%%%%%%%%%%%%
\subsection{Serious games}
Educative simulations are often in the form of a serious game. Anycare is a table-top role-playing game to involve stakeholders \cite{terti2019anycare}. LittoSIM \cite{becu2017participatory} is very realistic and aimed at emergency managers; SPRITE \cite{sprite2018} teaches risk management to engineering students; however both games focus on longer-term management and protection against coastal submersion (\eg building dykes) rather than communication. 

Many serious games are specifically targeted at children. It is argued that children are often more enthusiastic, motivated to learn, and receptive to new ideas \cite[p.142]{izadkhah2005towards}. Besides, they are a good channel to reach (and convince) their parents, and spread the ideas to the wider society \cite[p.19]{barreto2014thesis}. \cite[p.1]{fitzgerald2000education} confirm that it is better to develop an active (rather than fatalistic) mind-set about disaster risks at an ''early age'', and a culture of prevention takes time to form. However, engaging children requires engaging their teachers first.

In serious games, the player can take their own normal role, an imaginary role (invented for the game), or another existing role. This change of perspective is a powerful tool to get people to understand the specific point of view and challenges of a different role, and can make it easier to accept decisions made by that role. Shubik \cite{shubik1971} supports the usefulness of role playing a different position in order to understand another individual's point of view. Also of benefit in his view is the ability for participants to watch how stakeholders make their decisions in the game. This suggests that observing people playing can also be of interest, and that not only the players themselves learn from the game.

\subsection{Conclusion}
Existing simulators are most often aimed at training or supporting decisions of emergency managers. We adopt a different approach where we propose a serious game aimed at changing the population's perspective by letting them play the role of an emergency manager confronted with difficult decisions. This is in line with research showing that serious games are a good tool to provoke a change of perspective and improve mutual understanding. Our game will not only engage children (who are showed to be a good medium to reach the wider population) but also adults, their parents and teachers. 

This approach is therefore also in line with the risk communication principles advocating transparency and empowerment of the population \cite{reynolds2010principles}, a field currently lacking research. Our claim is that such a game will improve the population's awareness of the difficulty to predict and announce natural disasters, and of their own responsibility to react despite a certain rate of (unavoidable) false alarms. Section~\ref{sec:vigiflood} presents our game, before Section~\ref{sec:eval} describes the evaluation that we conducted to validate this claim.

%%%%%%%%%%%%%%%%%%%%%
%%% SECTION MODEL %%%
%%%%%%%%%%%%%%%%%%%%%

%%%%%%%%%%%%%%%%%%%%
\section{VigiFlood} \label{sec:vigiflood}
%%%%%%%%%%%%%%%%%%%%

The VigiFlood serious game is based on a conceptual agent-based model of human behaviour in flash floods. The validity of the underlying behaviour model is ensured by its grounding on psychological and sociological theories of trust and risk communication described above. The physical model of flood on the other hand need not be extremely realistic to reach an educational goal, in line with \cite{axelrod1997advancing}. This conceptual model has been described in \cite{iscram2019a}; the section provides a quick overview of the conceptual model and its implementation, before we proceed to describing its evaluation and concrete use, which are the core contributions of the current paper.

\subsection{Underlying conceptual agent model}

Our approach is to use an agent-based model of the population and their reaction to their environment (vigilance level, observed meteorological events). This section describes the agents representing the residents: their attributes, how they subjectively evaluate risk, how they update their trust, and how this subsequently impacts their behaviour.

\subsubsection{Attributes}
The population is composed of a number of heterogeneous resident agents, that have the same attributes but differ by their values:
\begin{itemize}
    \item Subjective risk (can be over-estimated or under-estimated compared to -inaccessible- objective risk)
    \item Risk aversion threshold (what level of risk they can tolerate before they choose to evacuate)
    \item Trust in vigilance messages (how much they believe the level of risk announced by the authorities)
    \item Memory depth/experience (how many past flood or vigilance events they remember)
    \item Risk evaluation strategy (how they evaluate risk based on these events, in line with psychological theories saying that humans evaluate risk based on previous experience and emotions \cite{cercpsy})
\end{itemize}

\subsubsection{Subjective risk evaluation}
Residents receive vigilance alerts, which indicate an 'official' level of risk, but not all residents equally believe in this level of risk. Indeed, each resident first subjectively evaluates risk (\ie expected rain) based on their \textbf{memory depth} and on their \textbf{risk evaluation strategy} which reflects their personality:
\begin{itemize}
    \item Optimistic (consider the minimum amount of rain observed during past events where the same vigilance colour was raised; if a false alarm was raised in the past this can lead to expect no or very little rain);
    \item Pessimistic (consider the maximum amount of rain; this is very forgiving to false alarms); 
    \item Rational (consider an average of observed rain on remembered past occurrences of the same vigilance level); 
    \item Short-memory (consider only the last occurrence; this simulates the loss of risk memory observed over time when no significant disaster happens \cite{fanta2019long}). 
\end{itemize}

\subsubsection{Interleaving with trust}
Trust and risk are interleaved:
\begin{itemize}
    \item Trust mediates subjective risk assessment: each resident's ponders their personal risk assessment vs the official communicated risk based on their (dynamic) level of \textbf{trust} in the vigilance alerts. Concretely, if trust is 100\%, the resident will fully trust the official risk and expect exactly the amount of rain corresponding to the vigilance colour; if it is 0\% they will fully trust their own judgement and expect exactly the subjective value resulting from their past experience.
    \item Residents then observe real rain and compare it with their expectations. If they are surprised by the actual amount of rain (either because it is higher or lower than expected / announced), their trust in the forecast will decrease (more if a flood was un-announced than in case of false alarm; and more for higher impact events). On the contrary if the observation is in line with expectations, trust will only slightly increase, as this is judged as being normal.
\end{itemize}

\subsubsection{Biased decision making}

Finally the residents' decision to evacuate early is based on the comparison of their subjective risk value with their personal risk aversion threshold. They can also evacuate after directly observing high amounts of rain (exceeding their aversion threshold), but this often happens too late, hence the importance of maintaining trust in pre-flood warnings.
Loss of trust pushes residents to neglect the official communicated risk, which can lead to two opposite situations:
\begin{itemize}
    \item After false alerts, personal risk assessment is low (memory is full of events with high vigilance but low rain); neglecting official risk conduces to under-estimating risk. Potential consequences can be serious: failure to prepare and evacuate in time.
    \item After missed alerts, personal subjective risk assessment is high (memory contains events with low vigilance but high rain); neglecting official risk conduces to over-estimating risk. Potential consequences include over-reacting, 
    % potential chaos, panic?
    which can be costly if it is not a resident but a stakeholder (mayor, etc) who takes unnecessary measures. %Besides, they will lose trust even more if they feel that they engaged costs 'for nothing'
\end{itemize}

%%%%%%%%%%%%%%%%%%%%%%%%%%%%%%%%%%%
%%% GAME INTERFACE AND GAMEPLAY %%%
%%%%%%%%%%%%%%%%%%%%%%%%%%%%%%%%%%%

\subsection{Game interface and gameplay}
The conceptual model described above was implemented in Python in the form of a serious game (the game presented here is a slight extension of \cite{iscram2019a}). The idea is that the player takes the role of risk communicator, having to decide the vigilance level (color between green, yellow, orange, red), based on uncertain clues (rain forecast). Their actions influence the population, whose trust and subjective risk level evolve over time, and who might or might not evacuate when floods are announced; they can also trigger various events simulating the reactions of institutional actors to the vigilance level (closing schools, stopping school bus services, closing roads, etc) and the impact of rain or floods on the environment (collapsed bridge, etc). The player can observe the impact of their actions through various information panels. The following paragraphs detail the interface of the game and its different phases.

\subsubsection{Interface}
\begin{center}
\begin{figure}
    \centering
    \includegraphics[scale=0.31]{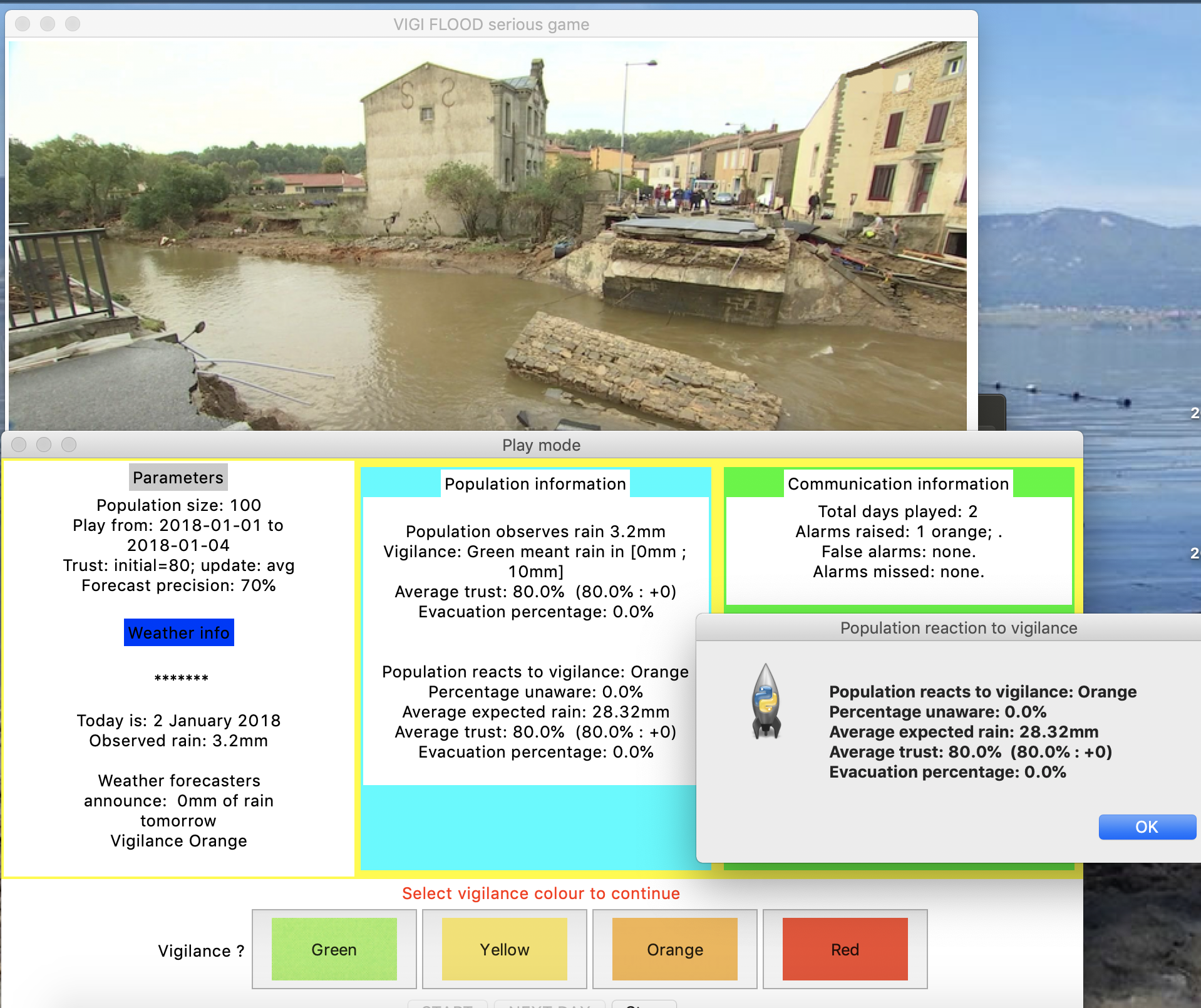}
    \caption{Screenshot of the game interface}
    \label{fig:gui}
\end{figure}
\end{center}

The interface (shown on Figure~\ref{fig:gui}) comprises several parts. 
\begin{itemize}
    \item Weather tab: reminds the date, the observed rain, the forecast for the next day (mm of rain), and the announced vigilance colour. 
    \item Population tab: details average subjective risk (expected rain) per vigilance colour, average trust in vigilance messages (with its last evolution), percentage of population that is unaware of risk (subjective risk is lower than objective risk), and evacuation percentage.
    \item Communication tab providing communication statistics: number of days played, and for each colour: number of vigilance days, number of false alarms (raised, but observed rain was lower than expected/ announced), number of alarms missed (not raised, and observed rain was higher than expected/ announced).
    \item Vigilance level selection buttons (green to red)
    \item Popups to display events triggered by the player's actions
\end{itemize}

\subsubsection{Game phases}
The game loop proceeds as follows: 
\begin{enumerate}
    \item The date of the day is displayed, as well as the observed amount of rain that day (in mm).
    \item The residents react to the observation of the rain: they compare it with the current vigilance level (set the day before) to update trust, and they might evacuate if needed.
    \item The player is provided with feedback about the population: average trust and its explanation (in terms of how much rain they expected based on the vigilance colour); and percentage of evacuated residents.
    \item The weather forecast service announces a forecast amount of rain for the next day.
    \item The player is asked to set a vigilance colour based on this forecast.
    \item The population reacts to this vigilance level: compute subjective expected rain based on previous similar alerts, and might evacuate if this is above their risk aversion threshold;
    \item The player receives feedback about the population's risk awareness percentage; their average expected rain (in mm); their average trust (in \%); and the percentage that evacuated as a result of the alert.
    \item Daily rain and vigilance are stored in game history, time moves forward to the next day.
\end{enumerate}

\subsection{Scenarios and data}

The game can be played either with a realistic scenario using actual rain and vigilance data, or with a generated pedagogical scenario that places them in specific intended situations aimed at testing their reactions. The advantage of the generated scenario is to accelerate the game and control the desired pedagogical sequence. Real data was extracted with Python scripts, from public meteorological archive websites.
\begin{itemize}
    \item Rain data from Infoclimat\footnote{Infoclimat: \url{https://www.infoclimat.fr}}, which provides archives of daily, monthly and yearly meteorological data (temperatures, wind, rain, etc) since 1973. We focused on Carcassonne-Salvaza, a meteorological station of the town of Carcassonne, the prefecture of the Aude department that was the most imapcted by October 2018 floods. We extracted the following data for the years 2010-2018: normal monthly rain (seasonal norms computed by Infoclimat between 1981-2010), number of days of rain (at least 1 mm) per month, and actual daily readings of rain amounts. 
    \item Vigilance data from Vigilance Public Meteo\footnote{\url{http://vigilance-public.meteo.fr/}} which provides maps and details of daily departmental vigilance alerts: time, level (green, yellow, orange, red), emitting agency (each covering a different region; Aude department depends on the CNP agency in Toulouse), and phenomenon concerned (floods, high winds, waves and coastal submersion, snow and black ice, etc). We extracted the daily rain and floods vigilance colours for the Aude department between 2010 and 2018 (green if no bulletin; higher colour if multiple ones).
\end{itemize}

%%%%%%%%%%%%%%%%%%%%%%%%%%%%%%%%%%%%%%%%%%%%%%%%%%%%%%%
%%% SECTION 4 - EVALUATION - SECTION 4 - EVALUATION %%%
%%%%%%%%%%%%%%%%%%%%%%%%%%%%%%%%%%%%%%%%%%%%%%%%%%%%%%%
%%% SECTION 4 - EVALUATION - SECTION 4 - EVALUATION %%%
%%%%%%%%%%%%%%%%%%%%%%%%%%%%%%%%%%%%%%%%%%%%%%%%%%%%%%%
%%% SECTION 4 - EVALUATION - SECTION 4 - EVALUATION %%%
%%%%%%%%%%%%%%%%%%%%%%%%%%%%%%%%%%%%%%%%%%%%%%%%%%%%%%%

%%%%%%%%%%%%%%%%%%%%%%%%%%%%%%%%%%
\section{Experimental evaluation} \label{sec:eval}

The game described above is designed to induce a change of perspective, where normal residents are put in a decider's shoes and faced with the responsibility to set the vigilance level themselves.
Our \textbf{hypothesis} is that this change of perspective should improve users' awareness of the difficulty to set vigilance without mistakes, and of their own responsibility for being vigilant. As a result, we also expect a shift of behavioural intentions towards more protective actions.
In order to test these hypotheses, we have designed a questionnaire simulating the game, and administered it online in order to reach a wide audience. 

The following paragraphs describe the questionnaire, the recruitment of participants, and the results along 2 axes: how the participants (subjectively) judged the game; and the (objective) impact that the game had on their awareness and intentions.

\subsection{Questionnaire}
The questionnaire is written in French. It was proof-read by a linguist, and validated by an ethics and data privacy consultant. It has also received the agreement of University Grenoble Alpes ethics committee\footnote{Agreement number CER Grenoble Alpes-Avis-2019-09-24-5}.
It is designed to proceed in the following six phases:
\begin{enumerate}
    \item Assess the responders' previous experience, knowledge of and trust in the French vigilance system.
    \item Assess the responders' awareness of risk, challenges, own responsibility; and behavioural intentions in case of floods (before).
    \item Change of perspective: exercise of setting vigilance level, in different more or less complex situations, with different clues.
    \item Subjective evaluation of the game by responders: interest, usefulness, willingness to use it.
    \item Re-assess the responders' awareness of risk, challenges, own responsibilities, and their behavioural intentions in case of floods (after).
    \item Demographic questions to categorise responders.
\end{enumerate}
The complete list of questions, translated into English, is available in Annex 1.

\subsection{Recruitment of participants}

We recruited adult participants by broadcasting the link to the questionnaire via email, through the author' professional and familial networks. The goal was to reach a wide audience, avoiding a classical bias of only testing software with computer scientists and students, and get answers from people with and without flood experience to enable comparisons. 

\begin{figure}[ht]
   \begin{minipage}[l]{.23\linewidth}
    \centering
    \includegraphics[scale=0.3]{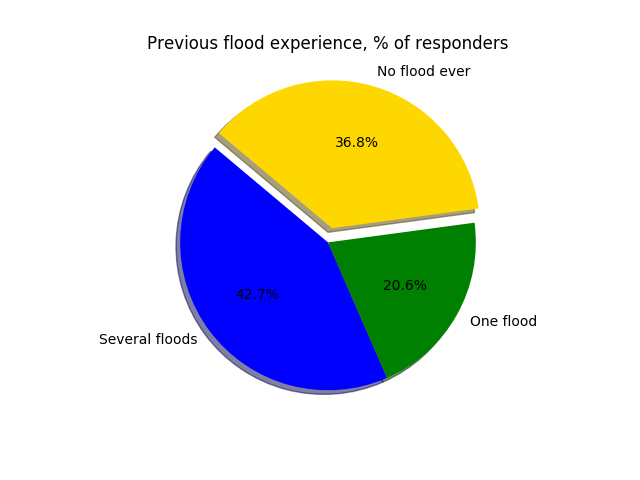}
    \caption{Experience} \label{pie1}
    \end{minipage} \hfill
   \begin{minipage}[l]{.23\linewidth}
    \centering
    \includegraphics[scale=0.3]{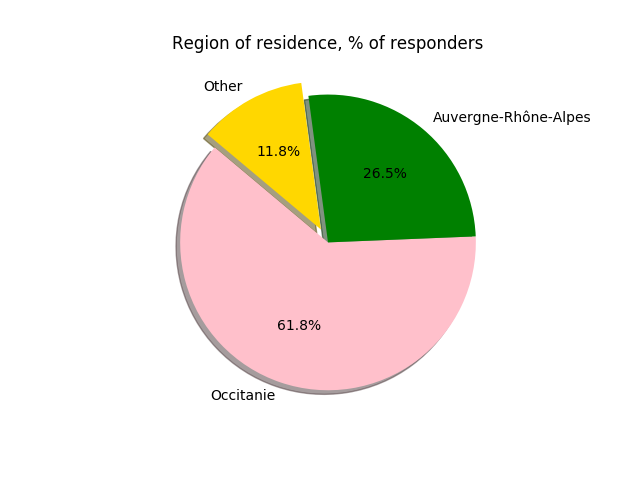}
    \caption{Residence} \label{pie2}
    \end{minipage} \hfill
   \begin{minipage}[l]{.23\linewidth}
    \centering
    \includegraphics[scale=0.3]{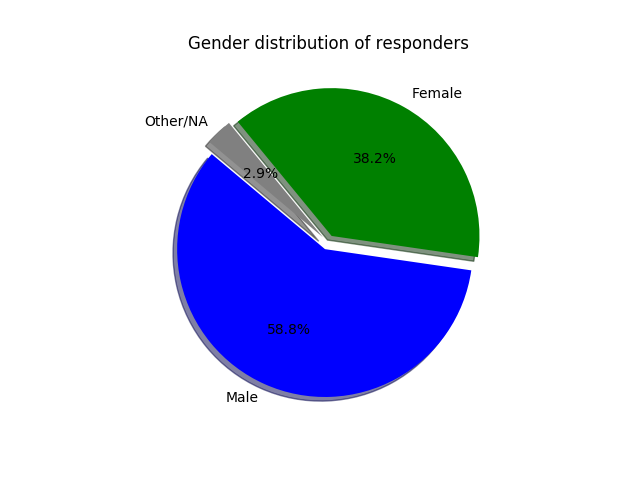}
    \caption{Gender} \label{pie"}
    \end{minipage} \hfill
   \begin{minipage}[l]{.23\linewidth}
    \centering
    \includegraphics[scale=0.3]{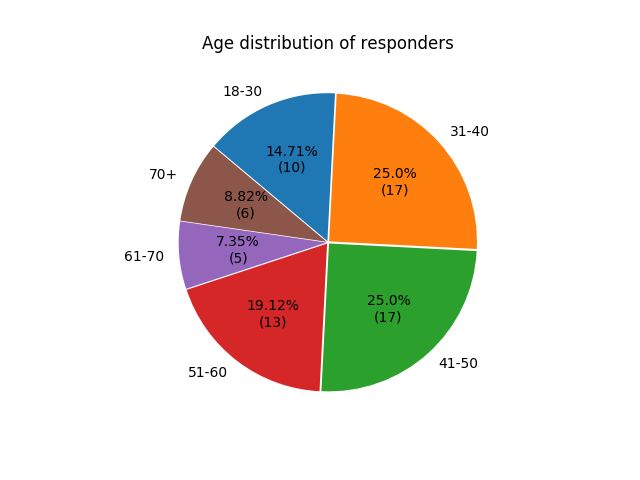}
    \caption{Age} \label{pie4}
    \end{minipage}
\end{figure}

The online questionnaire\footnote{Online questionnaire (readers are still welcome to answer): \url{https://forms.gle/XMSdxYCj3BGHrjX3A}} has received 80 answers so far. 
The pie charts in Figures~\ref{pie1} to \ref{pie4} show that our sample was rather representative with responders from all age groups, slightly more males than females, and having different levels of experience with floods. Due to our mode of recruitment, the responders mainly live in 2 regions: Occitanie (South-West region of France where the October 2018 floods occurred, reached via familial network) and Auvergne-Rhône-Alpes (South-East region of France where the author works, reached via professional network).

\subsection{Results: subjective evaluation of the game}
The responders were asked to score VigiFlood (in its simulated version administered via the online questionnaire) on 5 criteria:
\begin{itemize}
    \item Interest: how interesting was it to answer or play?
    \item Boredom: how boring is the current gameplay?
    \item Usefulness: how useful is it to learn about floods, forecast and communication?
    \item Town training: how willing that such a game would be offered as training by the town council?
    \item School training: how willing that such a game would be offered to children as part of school programs about natural hazards?
\end{itemize}

\begin{center}
\begin{figure}[ht]
    \centering
    \includegraphics[scale=0.6]{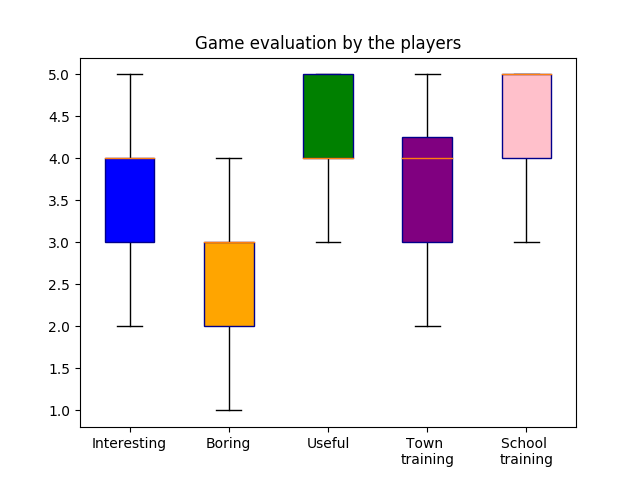}
    \caption{Evaluation of the game (5-point scale scores): interest, boredom, usefulness, willingness to use it for training of town residents (by town council), or for training of school children (by professors).}
    \label{fig4:scores}
\end{figure}
\end{center}

Figure~\ref{fig4:scores} summarises the results in the form of a boxplot diagram: the boxes extend from lower to upper quartile values of the scores, with a line at the median; the whiskers extend from the boxes to show the range of the data; outliers are not shown here.

We can see that responders find the game very useful (avg= 4.18; stdev= 0.98). They are willing to have it offered by their town council as training (avg= 3.79; stdev= 1.08), and even more willing to see it offered to children as part of school programs about natural hazards (avg= 4.35; stdev= 0.84). However, even though they found the concept of the game quite interesting (avg= 3.65; stdev= 1.04), they also judge the gameplay rather boring (avg= 2.51; stdev= 0.95). Future work is needed to improve immersion and engagement with the game. We will also further analyse if these scores are different between people with or without flood experience.

\subsection{Results: objective impact of the game}
Another goal of this questionnaire was to assess how playing the game (or its simulated version) can influence awareness. We measured a number of awareness criteria (risk awareness, awareness of own responsibility, intended actions in case of orange or red vigilance) before (phase 2) and after (phase 5) the change of perspective induced by the core of the questionnaire (phase 3). 

\begin{figure}[ht]
   \begin{minipage}[l]{.46\linewidth}
      \includegraphics[scale=0.5]{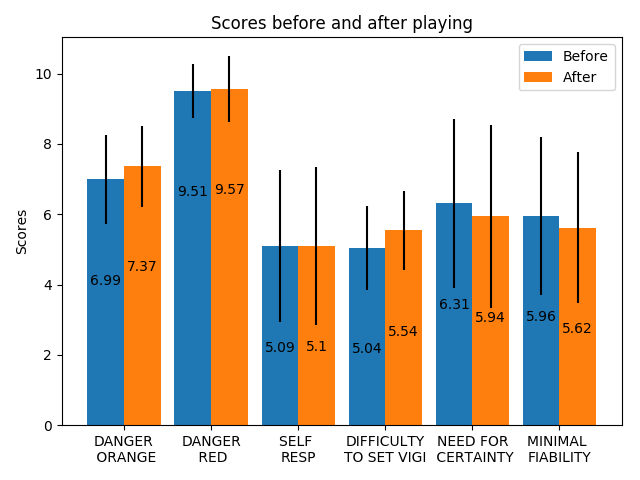}
        \caption{Evolution of awareness scores before/after playing}
    \label{fig1:awareness}
   \end{minipage} \hfill
   \begin{minipage}[c]{.46\linewidth}
     \includegraphics[scale=0.5]{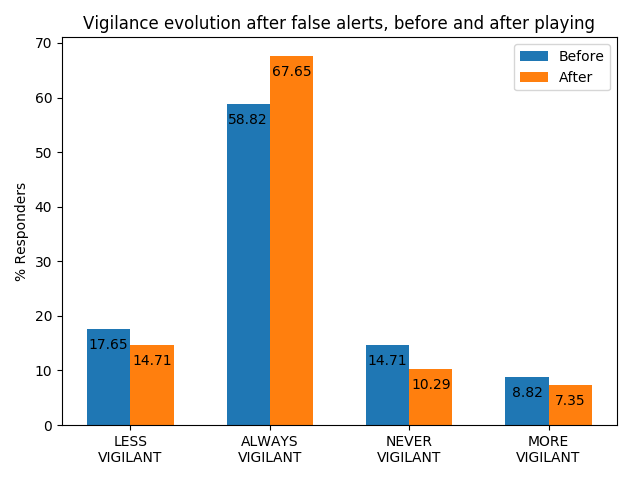}
    \caption{Evolution of loss of vigilance with false alerts, before/after playing}
    \label{fig2:vigilance}
   \end{minipage}
\end{figure}

\subsubsection{Impact on awareness}
Figure~\ref{fig1:awareness} shows how awareness scores evolved between before and after the role-playing sequence. We can see that risk awareness increases for orange vigilance (that tend to be ignored when too frequent), which is a good thing. It does not increase for red vigilance, which are probably already alarming enough. The awareness of own responsibility in self protection does not change significantly after the game, probably because the focus was more on the alert phase, and the game does not show actions of the population; further developments of the software will try to focus more on the population side. Finally awareness of the difficulty of setting the right level of vigilance does increase, which we hope should limit the loss of trust induced when the vigilance colour is perceived as wrong.

\subsubsection{Impact of 'false alerts' on trust}
Figure~\ref{fig2:vigilance} illustrates the dynamics of population vigilance over multiple alerts. It shows that before playing (blue bars) 17.65\% of responders report being less vigilant after what they perceive as a "false alert", while 14.71\% report never being vigilant anyway. 

This is worrying for several reasons. First, ''false alarms'' can be quite frequent, because it is hard to predict such phenomenons, and because the vigilance is set at the entire department level. Therefore even if a flood does indeed happen locally, it is possible that a large part of the department is spared and left to believe that it was a false alarm. Second, even if it is actually a false alarm and no flood happens this time, intense rain might leave the soils saturated and river levels very high, thus favouring future floods if more rain happens later. As a result, past alerts should increase vigilance rather than decreasing it.

On the figure we can see that after playing (orange bars), the percentage of responders reporting that they would be less vigilant or never vigilant have both dropped, while the percentage of responders intending to be always vigilant has risen from 58.82\% to 67.65\%.

The survey was administered to many people having recently lived the October 2018 floods, which explains why most users in our sample are already very vigilant. Further analysis is required to compared the dynamics of vigilance in responders with or without flood experience. It would also be interesting to compare with people having a less recent flood experience, as the literature reports how humans tend to cyclically forget about risk until a new crisis occurs.

\subsubsection{Protective measures taken}
\begin{center}
\begin{figure}[ht]
    \centering
    \includegraphics[scale=0.7]{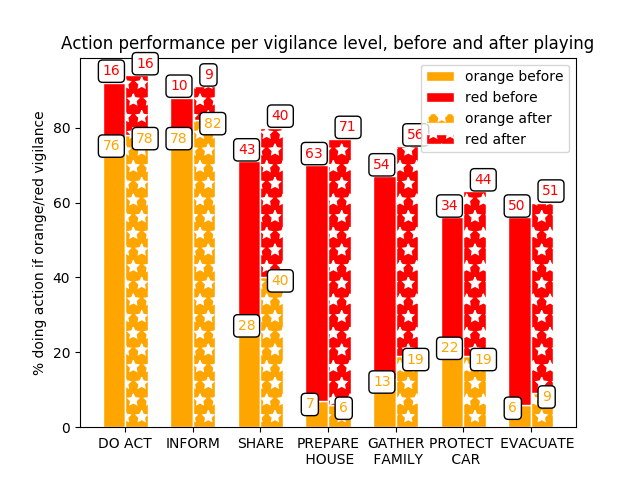}
    \caption{Evolution of actions performed/intended if orange/red vigilance, before/after playing}
    \label{fig3:actions}
\end{figure}
\end{center}

Figure~\ref{fig3:actions} illustrates the percentage of responders taking various protective measures, before (questions in phase 2 of the questionnaire) vs after playing (same questions in phase 5 of the questionnaire). The actions mentioned in the survey are: search information (INFORM); share information (SHARE); prepare house; gather with relatives and family (GATHER); park one's car on higher ground (PROTECT CAR); evacuate; or do nothing special. In the figure, each bar shows in orange the percentage of responders saying they would take this action when an orange vigilance is announced, and in red at the top the additional percentage of responders saying they would take this action only when a red vigilance is announced. The left part of each bar is before the game, while the right part is after. 

The left bar measures the percentage of responders doing ``something'' at all. We can see that about a quarter of the population would do nothing at all in case of orange vigilance (24\% before the game, 22\% after); very few would do nothing if there was a red vigilance (8\% before the game, 6\% after). This high level of behavioural intentions is probably due to our sample comprising more people with flood experience than the general population.

The most frequently intended action is to search for information: 78\% will try to get more information in case of an orange vigilance, and an additional 10\% (so a total of 88\%) in case of a red vigilance. These percentages rise to 82\% and 91\% after playing the game. We see a similar impact on the actions of sharing information, gathering with family, or evacuating (which is mostly envisaged in case of red vigilance). On the other hand, intentions for preparation actions (preparing house and protecting car) decrease slightly in case of orange vigilance, but increase in case of red vigilance. 

This hence shows that changing perspective through the questionnaire has modified the intentions of the responders. However, these are only declared intentions, they might be influenced by a willingness to give the ``right'' or expected answer; they might also be actual intentions but that do not lead to actions once under stress in a real flood situation. In any case, it at least shows an improvement after the game of the awareness of actions that should or should not be performed.

\subsection{Conclusion of the evaluation and future work} 
This survey was intended to show the potential of such a serious game for getting the population of flood-prone areas to change perspective. Since the software is still under development, and to allow for a wider participation in different regions, it was decided to run the survey online via a questionnaire that simulates the intended functioning of the Vigiflood game. This is a first limitation, since we evaluated a slightly different process than that of the actual game. However, we believe that both versions do induce the same change of perspective, which is the key element of our approach.

The survey measured the impact of this role-playing exercise on the users' awareness, trust, and protective actions intentions. Our results show a positive impact on these indicators, which is encouraging. Another limitation of this survey comes from our sample of responders. Only people who felt concerned with floods and motivated by the topic would take the time to answer the survey, given its repetitiveness and lack of engaging features. We hope that the final software, once we make it less boring and more immersive, will do even better in allowing the users to change perspective and to gain insight about flood risk communication. The gameplay design and implementation will be subject to future work, before we can use this serious game with high school students. 

Further analysis will also be performed on the answers, for instance to categorise the users' risk evaluation strategies (what strategy do they use to choose a vigilance level), or to compare answers of different profiles of responders (with or without flood experience; by age or gender; etc). This should lead to a better understanding of the population's subjective risk analysis and the factors influencing it.

%%%%%%%%%%%%%%%%%%%%%%%%%%%%%%%%%
%%% CONCLUSION %%% CONCLUSION %%%
%%%%%%%%%%%%%%%%%%%%%%%%%%%%%%%%%
\section{Conclusion and prospects}  \label{sec:cci}

In this paper we presented Vigiflood, a serious game for raising awareness about the challenges of flood risk communication. In relies on an agent-based model of the population's trust and decisions, itself grounded in psychology and sociology of human behaviour. A first prototype of the game is implemented and functional. It lets the player choose the flood vigilance colour (green, yellow, orange, red) based on the weather forecast (generated from real meteorological data extracted from archives for the target area). It is therefore a role-playing game where the users (residents in flood prone areas) take on a different role in the game (that of weather forecasters) than in real life. The main idea behind this game design is that the change of perspective induced by playing a different role will lead to a better understanding of the difficulties of this role, specifically here the challenges of announcing the ``right'' level of vigilance.

The main contribution of this paper is the evaluation of the concept of the game. We ran an online survey replicating the role-playing part of the game, preceded and followed by questions evaluating the user's awareness of risks, understanding of difficulties, vigilance, and behavioural intentions in case of flood risk. This evaluation shows encouraging results, where the change of perspective does induce better awareness of risks, more protective actions intentions, and better vigilance and trust in the forecast. However, only very motivated users would answer the survey or play with the prototype, due to its boring and repetitive design. More work is needed to improve the game design and playability, in order to allow Vigiflood to be played by the general public, and in particular by high school students. This is essential to guarantee that this serious game can have an actual beneficial impact during future flood events.

The underlying population model can also be turned into an interactive simulator to train weather forecasters to take the subjective reactions of the population into account when announcing a forecast. This alternative version of Vigiflood is also under work. At a time when more and more floods are expected to happen in some parts of the world \cite{roudier2016projections,kerr2007global,schiermeier2011increased}, we believe that agent-based simulation can provide very useful tools to train and educate both the population and the deciders in order to reduce impact of these floods.

\section*{Acknowledgements}

This work is supported by the French National Research Agency in the framework of the Investissements d’Avenir program (ANR-15-IDEX-02), project Risk@UGA.

\footnotesize
\bibliographystyle{plain}

\newpage
\section*{Annex1: list of questions (translated from French)}

\footnotesize
\begin{itemize}
    \item Phase 1: experience, awareness and trust
    \begin{enumerate}
        \item Have you experienced floods before?
        \item Did these floods affect your residence?
        \item Were these floods announced in advance?
        \item Did these floods require your evacuation?
        \item Do you know about MeteoFrance vigilance system? 
        \item Do you know about Vigicrues monitoring website? 
        \item Do you trust these meteorological forecast in case of floods?
        \item Do you trust the local authorities to warn and protect you in case of floods?
    \end{enumerate}
    \item Phase 2: awareness of risk, challenges, and self responsibility; intended actions (BEFORE)
    \begin{enumerate}[resume]
        \item Who do you think is responsible for your information and protection in case of floods?
        \item How easy/hard do you think it is to announce the right level of vigilance?
        \item What information do you think the authorities rely on to set the vigilance level?
        \item How important is it to be certain of a forecast before announcing a vigilance level to the population?
        \item What is the minimal reliability required before announcing a level of vigilance?
        \item In your opinion, what level of danger does an orange flood vigilance indicate?
        \item In your opinion, what level of danger does a red flood vigilance indicate?
        \item In case of orange/red flood vigilance, what do you do? (check all actions among: search information, share information, prepare house, gather with relatives, park car higher, evacuate, nothing special)
        \item During the last 2 orange vigilances, there were no flood in the end. What is your reaction to the next orange vigilance? (trust it anyway, distrust it anyway, trust it more, trust it less)
\end{enumerate}
    \item Phase 3: change of perspective, setting vigilance in different situations:
    \begin{enumerate}[resume]
        \item Series of questions of the form: the weather forecast services announce  XXX mm of rain with a confidence index of XXX \%. Which vigilance colour do you wish to announce?
        \item Series of similar questions with additional contextual elements
        \item Series of similar questions with additional information about previous false alarms and population trust
    \end{enumerate}
    \item Phase 4: evaluating the game (rate on a 5-point scale):
    \begin{enumerate}[resume]
        \item Such a game would be interesting to play
        \item Such a game would be boring and repetitive
        \item Such a game would be useful to learn and understand the vigilance system
        \item You would like your town council to offer such a game as part of its flood risk prevention plan
        \item You would like schools to offer such a game to children as part of their program about natural hazards
    \end{enumerate}
    \item Phase 5: re-assessing awareness and intentions (AFTER):
    \begin{enumerate}[resume]
        \item Same questions as in phase 2
        \item Have you discovered new criteria for the definition of the flood vigilance level that you had not considered before? If yes, which ones?
    \item Free comments
    \end{enumerate}
    \item Phase 6: demographic questions: region of residence, age, gender, job (if related with weather forecast or depending on weather conditions)
\end{itemize}

\end{document}